\documentclass{Interspeech2024}
\usepackage[T1]{fontenc}
\usepackage[utf8]{inputenc}

\usepackage{caption}
\usepackage{subcaption}
\usepackage{amssymb}
\usepackage{wrapfig}

\usepackage{eqparbox}
\usepackage{float}
\usepackage{lipsum}
\usepackage{multirow}
\usepackage{scalerel}
\usepackage{hyperref}
\usepackage{amsmath}
\usepackage{mathtools}
\usepackage[capitalize]{cleveref}
\Crefname{chapter}{Chap.}{Chaps.}
\Crefname{section}{Sec.}{Secs.}
\Crefname{figure}{Fig.}{Figs.}

\def\am{$\theta_{\scaleto{\text{AM }}{4pt}}$}
\def\lm{$\theta_{\scaleto{\text{LM}}{4pt}}$}

\newcommand{\argmax}{\operatornamewithlimits{argmax}}
\newcommand\numberthis{\addtocounter{equation}{1}\tag{\theequation}}

\def\L{{\cal L}}
\makeatletter
\renewcommand{\section}{\@startsection
	{section}%
	{1}%
	{}%
	{-0.7\baselineskip}%
	{0.1\baselineskip}%
	{}}%
\renewcommand{\subsection}{\@startsection
	{subsection}%
	{2}%
	{}%
	{-0.3\baselineskip}%
	{0.2\baselineskip}%
	{}}%
\renewcommand{\subsubsection}{\@startsection
	{subsubsection}%
	{3}%
	{}%
	{-0.1\baselineskip}%
	{0.1\baselineskip}%
	{}}%
\g@addto@macro\normalsize{%
	\setlength\abovedisplayskip{3pt plus 0pt minus 1pt}
	\setlength\belowdisplayskip{3pt plus01pt minus 1pt}
	\setlength\abovedisplayshortskip{3pt plus 1pt minus 1pt}
	\setlength\belowdisplayshortskip{3pt plus 1pt minus 1pt}
}
\makeatother

\usepackage{xcolor}

\interspeechcameraready

\title{Investigating the Effect of Label Topology and Training Criterion\\ on ASR Performance and Alignment Quality}

\name[affiliation={1}]{Tina}{Raissi}
\name[affiliation={2}]{Christoph}{L\"uscher$^\star$}
\name[affiliation={1,2}]{Simon}{Berger$^\star$}
\name[affiliation={1,2}]{Ralf}{Schl\"uter}
\name[affiliation={1,2}]{Hermann}{Ney}

\address{ 
  $^1$Machine Learning and Human Language Technology Group, RWTH Aachen University, Germany\\
  $^2$AppTek GmbH, Germany\\
{\normalsize $^\star$ Denotes equal contribution}}
\email{\{raissi,schlueter\}@cs.rwth-aachen.de}


\begin{document}

\maketitle

\begin{abstract} 
The ongoing research scenario for automatic speech recognition~(ASR) envisions a clear division between end-to-end approaches and classic modular systems.\ Even though a high-level comparison between the two approaches in terms of their requirements and (dis)advantages is commonly addressed, a closer comparison under similar conditions is not readily available in the literature.\ In this work, we present a comparison focused on the label topology and training criterion.\ We compare two discriminative alignment models with hidden Markov model~(HMM) and connectionist temporal classification topology, and two first-order label context ASR models utilizing factored~HMM and strictly monotonic recurrent neural network transducer, respectively.\ We use different measurements for the evaluation of the alignment quality, and compare word error rate and real time factor of our best systems. Experiments are conducted on the LibriSpeech 960h and Switchboard 300h tasks. 
\end{abstract}

\section{Introduction}
 According to the latest version of the survey \textit{Progress and Prospects for Spoken Language Technology}, there is a general consensus that by the year 2025 the majority of automatic speech recognition~(ASR) systems have completely abandoned the hidden Markov model~(HMM) paradigm~\cite{moore2023progress}.\ Even though rich relations exist between HMM and current approaches often termed as end-to-end~\cite{prabhavalkar2023end}.\ In addition to the first-order Markov assumption, HMM builds upon a conditional independence assumption.\ According to this assumption, when conditioned on a state, the aligned frames within that segment are independent of each other, as well as of all the frames generated by the other states.\ Nearly all HMM based alignment systems used in literature are trained by maximizing the likelihood of the training data with respect to a set of Gaussian mixture model parameters.\ A generative model such as GMM needs to cover for all possible observation sequences and therefore requires that the input sequence is restricted to a local representation with no access to wider ranges of dependencies.\
In the years following the success of large vocabulary ASR with hybrid-HMM, substantial research efforts have been directed towards addressing the inherent simplifying assumptions in hidden Markov approach.\ This has been done by considering a wider range of dependencies, applying global normalization, and performing sequence discriminative training~\cite{woodland2002large,berger1996maximum,lafferty2001conditional,gunawardana2005hidden}.\ In the recent years, there has been exploration into the simplification of the standard HMM pipeline by eliminating GMM and HMM state-tying~\cite{hadian2018end,raissi2023competitive}.\ Moreover, a simulation study verifying the effect of different simplifying assumptions of classic HMM has notably underlined the significant influence of the aforementioned output independence assumption on the ASR performance~\cite{gillick2011don}.\ The recent encoder architectures are equipped with the self-attention mechanism and recurrent layers which have the capability to capture
global dependencies of the input.\ One question to consider is whether utilizing downsampling over the time axis within these long-context neural networks allows the creation of an input representation that is more in line with the HMM independence assumption.

The current research scenario seems to have a preference towards the end-to-end approaches, primarily due to their simplicity and the widespread accessibility of resources and frameworks.\ However, when comparing the ASR performance of these approaches with classic HMM-based systems, various issues arise: differences in speech representation and frame shifts on the front-end side, as well as variations in label units, amount of data, use of external resources such as lexicon, alignment, and language model~\cite{rouhe2023principled,gimeno2024comparison}.\

In this work, we propose a comparison between two pairs of time synchronous models, by fixing all the conditions mentioned above, and considering different label topologies and training criteria.\ The specific choice of the models derives from a more high-level comparison between a purely discriminative sequence-to-sequence approach and a factored hybrid HMM system.\ The pipeline for both systems starts with training a zero-order label context discriminative alignment model from-scratch.\ Subsequently, after conducting forced alignment, a first-order label context model is trained for each system.\ As a further investigation step, we analyze the quality of the alignments for each topology using different measurements, and show improvements with respect to both word error rate~(WER) and real time factor~(RTF) for factored hybrid HMM system with a 40ms frame shift.\ All experiments are verified on both LibriSpeech 960h and Switchboard 300h.
\section{Modeling Approach}
\label{sec:modelapproach}
For an acoustic feature sequence X of length $T^{\prime}$, and its corresponding word sequence W of length N, let $h_1^{T} = Enc(X)$ denote the output of a neural encoder. We consider $T = \lceil T^{\prime}/m \rceil$ for m $\in \{1, 4\}$, where m = 4 realizes the common downsampling technique used within the neural encoder architectures.\ The inference rule for any ASR task is Bayes decision rule~\cite{bayes1763lii}, which maximizes the a-posteriori probability of a word sequence W given the input sequence, as described in \cref{eq:bayes1}.\ In the classic approach, an equivalent reformulation via Bayes identity in terms of separate acoustic and language models is carried out.\ The resulting generative approach having distinct parameters \am and \lm, respectively, is shown in \cref{eq:bayes2}.\ 
\begin{equation}\footnotesize
	\label{eq:bayes1}
	X \rightarrow\tilde{W}(X) = \underset{W}{\argmax} \left\lbrace P_{_{\theta}}(W | X)  \right\rbrace \phantom{0000000000}\vspace{-0.2cm}
\end{equation}
\begin{equation}\footnotesize
	\label{eq:bayes2}
	\phantom{X \rightarrow\tilde{W}(X)}= \underset{W}{\argmax} \left\lbrace P_{_{\theta_{\scaleto{\text{AM}}{2pt}}}} (X | W) \cdot P_{_{\theta_{\scaleto{\text{LM}}{2pt}}}}(W) \right\rbrace \vspace{-0.2cm}
\end{equation}

\subsection{Unifying Principles}
\label{subsec:principles}
In order to verify the effect of the label topology and the training criterion across different models, we fix some of the basic conditions.\ The underlying speech representation adheres to the classic HCLG~\cite{mohri2002weighted} composition unfolded over time, i.e. an alignment finite state acceptor~(FSA) $\mathcal{A}$ that defines all valid alignment paths between the input speech and the output label sequence.\ Independent from the modeling approach, the word sequence is generally mapped to a sequence of sub-word units.\ Let the sequence $\phi$ of length $M$ be the phonetic output label sequence of $W$, and disregard pronunciation variants.\ We augment each phoneme with its neighboring left and right phonemes for the classic triphone structure of the C composition in HCLG.\ Unfolding the result of the composition with a specific label topology over time is equivalent to a marginalization over a hidden alignment variable.\ In practice it gives the FSA $\mathcal{A}$ that contains all the allowed paths within a certain topology.\ The training criteria used for our models belong to the cross-entropy based family of training objectives.\ The criterion considered on the sequence level uses the sum over $\mathcal{A}$, and on the frame level restricts the loss to a single best path.\ 
With single state phonemes as label units, we explore different label topologies and probability distribution models for the frame-wise acoustic scores, as later explained in \cref{subsec:framework}.\ The label set for all topologies consists of the phonemes from the lexicon with the end-of-word (EOW) distinction~\cite{zhou:phoneme-transducer:2021}.\ For CTC and transducer an additional blank token is included and only HMM uses an explicit silence token.\ We use a downsampling of factor four on the ten milliseconds~(ms) input features for a total frame shift of 40ms.\

\subsection{Mapping of the Alignment Path}
\label{subsec:alignmap}
We showed how a word sequence can be mapped to its alignment FSA $\mathcal{A}$.\ For the inverse mapping within our proposed notation, we introduce two auxiliary functions: (1) for the blank-based topologies, denote $Y$ as a blank-augmented alignment sequence of length $T$, we use a function $\mathcal{B}$ which eliminates all blanks for an alignment path within $\mathcal{A}$ and returns the corresponding phoneme sequence, and (2) for the HMM topology, by using the hidden state sequence $S$ of length $T$, we denote $a_{n_{s_t}}$ as a function that takes as input the identity of the aligned state at time frame $t$ within a phoneme of the word at position $n$ and returns the phoneme label.\

\subsection{Conceptual Framework}
\label{subsec:framework}

Without using the WFST formalism, we describe each approach via a three-fold conceptual framework. Denote $\mathcal{F} = \{\Sigma, \mathcal{Q}, \mathcal{C}\}$ to be a tuple consisting of a label topology $\Sigma$, a frame-wise probability distribution model $ \mathcal{Q}$, and a training criterion $\mathcal{C}$.\\ 
\textbf{$\bullet$ Label Topology $\mathbf{\Sigma}$:} describes a set of rules related to the emission of a label out of the set of labels, at each time step.\ A label topology together with the label unit determine the set of all allowed paths within the alignment lattice $\mathcal{A}$.\\
\textbf{$\bullet$ Frame-level probability distribution model $\mathbf{\mathcal{Q}}$:} is obtained via application of model-specific assumptions for the decomposition of the sequence-level probability distributions of \cref{eq:bayes1,eq:bayes2} into frame-level probability models.\ The decomposition step generally follows a marginalization over a hidden alignment sequence variable.\  In the classic FST framework, {$\mathbf{\mathcal{Q}}$ can be simplified to the type of acoustic weight associated with each arc.\ We model $n$-order label context acoustic models for $n \in \{0,1\}$ with local normalization, i.e. with normalized probability distribution over the set of all labels at each time frame.\ At each time frame, we model posterior probabilities conditioned on the encoder output for both orders and conditioned additionally on the previous acoustic label context for the first order models.\\
\textbf{$\bullet$ Training Criterion $\mathbf{\mathcal{C}}$:} expresses the optimization objective.\ While all considered criteria can be viewed as a form of cross-entropy, there are differences caused by the specific modeling assumptions. For instance, we consider a sequence-level conditional likelihood with sum over all alignment paths, denoted as full-sum criterion, and a frame-wise cross-entropy or Viterbi training using the single best path.

\subsection{Models}
\label{subsec:models}

\subsubsection{Zero-Order Label Context Alignment Models}
\label{subsub:zeroorder}
We examine two discriminative alignment models that differ by label topology and frame level probability distribution.\ Both models are trained from scratch with conditional log-likelihood criterion.\ One model uses the classic CTC label topology with blank and label loop~\cite{graves2006connectionist}.\ The second model, proposed previously as posterior-HMM~(P-HMM)~\cite{Raissi+Zhou+:2023}, consists in an HMM topology with loop and forward transitions only.\
We start with the marginalization of the acoustic model with the hidden alignment variable, and we further insert the label topology constraint and model-specific assumptions while decomposing the sequence-level posterior into frame-level probability distributions.\ The definition of $\mathcal{Q}$ for CTC in \cref{eq:ctc} and for P-HMM in \cref{eq:HMM} differs by the transition probability $P(s_t| s_{t-1})$.\ Both models use a zero-order label posterior.\ However, the HMM uses an explicit transition probability $P(s_t| s_{t-1})$ normalized over loop and forward probability.\ 
\begin{equation}\label{eq:ctc} \scriptsize
		\hspace{-1.6cm}\sum_{Y} P(\phi, Y | h_1^{T}) =  \sum_{y_1^{T}:\phi_1^M} \prod_{t=1}^T P(y_t | h_t)\vspace{-0.3cm}
\end{equation}
\begin{equation}\scriptsize \label{eq:HMM}
	\sum_{S} P(\phi, S | h_1^{T}) =\sum_{s_1^T:\phi_1^M} \prod_{t=1}^{T} P(a_{n_{s_t}}| h_t) P(s_t| s_{t-1}) 
\end{equation}
The transition probability is generally omitted in case of CTC and given implicitly by the alignment sequence, unless considered within the Auto Segmentation Criterion~\cite{collobert2016wav2letter}.\ Similar assumptions can be done also for the simple variant of the P-HMM where loop and forward probabilities may be set to $0.5$ for all labels, and therefore they may be dropped, being constant with respect to both the training criterion and the inference rule~\cite{Raissi+Zhou+:2023}.\ Training is carried out for both models with conditional log-likelihood described in \cref{eq:fsloss} as follows: \vspace{-0.1cm}
\begin{equation}
	\label{eq:fsloss}\footnotesize
	\L = -\log \hspace{1mm} P(\phi| h_1^{T})  
\end{equation}
In practice, the two models learn the parameters $\theta$ for the inference rule in \cref{eq:bayes1} via two different alignment FSAs $\mathcal{A}$ and using two different probability distributions $\mathcal{Q}$.\ For P-HMM, the inference rule, shown in \cref{eq:zeroorderdecode}, uses the label and transition probabilities with scales $\alpha$, and $\beta$, respectively, combined with an external language model~(LM) scaled by $\lambda$.\ Moreover, a common internal language model~(ILM) subtraction with scale $\gamma$ is applied.\ Decoding for CTC follows a similar rule without using the transition model. \vspace{-0.2cm}
\begin{equation}	 \footnotesize
\label{eq:zeroorderdecode}
	\hspace{-0.3cm} \underset{W}{\argmax} \hspace{1mm}  \left\lbrace P^{\lambda}_{\text{LM}}(W) \underset{s_1^T:\phi_1^M:W}{\max} \prod_{t=1}^T \frac{P^{\alpha}(a_{n_{s_t}} | h_t)}{P^{\gamma}_{\text{ILM}}(a_{n_{s_t}})} P^{\beta}(s_t| s_{t-1})  \right\rbrace 
\end{equation} 
\subsubsection{First-Order Label Context ASR Models}
\label{subsub:firstorder}
The two first-order label context models are also built upon a common training criterion.\ However, in addition to different label topologies and frame level probability distributions, the models are designed to provide the acoustic parameters for different inference rules.\ 
Regarding the label topology $\Sigma$, we compare a strictly monotonic RNN-T topology~(mRNN-T)~\cite{tripathi2019monotonic,Zhou+Michel+:2022} and a context-dependent HMM without state-tying presented as factored hybrid~(FH)~\cite{raissi2020fh}.\ Both models use a frame-wise cross-entropy loss.\ We first discuss the $\mathcal{Q}$ component and the decision rule for the factored hybrid HMM case.\ The starting point in this approach is shown in \cref{eq:bayes2} and the model we consider provides \am, i.e. the parameters of the acoustic model in the generative form.\ The formulation in \cref{eq:first-order-HMM} defines a generative first-order label context (diphone) model, consisting of two separate factors, together with a context-dependent diphone state prior.\ \vspace{-0.1cm}
\begin{equation}\scriptsize
	 \sum_{S} P(h_1^{T}, S | \phi )= \hspace{-2mm} \sum_{s_1^T:\phi_1^M} \prod_{t=1}^{T} P(h_t | a_{n_{s_t}}, a_{n_{s_t}-1}) P(s_t| s_{t-1}) \nonumber
	\end{equation}\vspace{-0.2cm}
	 \begin{equation}\scriptsize
	\hspace{-2mm}= \hspace{-2mm} \sum_{s_1^T:\phi_1^M} \prod_{t=1}^{T}\frac{P(a_{n_{s_t}}| a_{n_{s_t}-1},  h_t) P(a_{n_{s_t} - 1} |  h_t) P(h_t)}{P_{\scaleto{\text{Prior}}{3pt}}(a_{n_{s_t}}, a_{n_{s_t}-1})}  P(s_t| s_{t-1}) 	\numberthis \label{eq:first-order-HMM}
\end{equation}
 The decision rule for FH in \cref{eq:firstorder-hmm-decode} uses both the language model and the prior as part of the decomposition and model definition.\  
\begin{equation} \scriptsize
	\hspace{-2mm}   \underset{W}{\argmax} \hspace{-0.5mm}  \left\lbrace \hspace{-0.5mm}P^{\lambda}_{\text{\lm}}(W)  \underset{\scaleto{s_1^T:\phi_1^M:W}{6pt}}{\max} \prod_{t=1}^T \hspace{-1mm} \frac{P(a_{n_{s_t}}, a_{n_{s_t}- 1})| h_t)}{P^{\gamma}_{\scaleto{\text{Prior}}{3pt}}(a_{n_{s_t}}, a_{n_{s_t}-1})} P^{\beta}(s_t| s_{t-1})  \right\rbrace  \numberthis \label{eq:firstorder-hmm-decode} 
\end{equation} 
The decomposition for the label posterior in mRNN-T includes only one of the factors from \cref{eq:first-order-HMM}, due to the direct discriminative approach:\vspace{-0.3cm}
\begin{equation}\scriptsize
	\sum_{Y} P(\phi, Y | h_1^{T})=\sum_{y_1^T:\phi_1^M} \prod_{t=1}^{T} P(y_t| a_{n_{y_{t-1}}},  h_t)	\numberthis \label{eq:first-order-rnnt}
\end{equation}
Note that here we overload the $a_n$ function to accept a blank augmented alignment label.\ The language model used in \cref{eq:firstorder-rnnt-decode} should be considered as an external LM and we subtract the internal LM instead of the context-dependent label prior.
\vspace{-0.4cm}
\begin{equation}\scriptsize
	\underset{W}{\argmax} \hspace{1mm}  \left\lbrace P^{\lambda}_{\text{LM}}(W) \underset{\scaleto{y_1^T:\phi_1^M:W}{6pt}}{\max} \prod_{t=1}^T \frac{P(y_t | a_{n_{y_{t-1}}}, h_t )}{P^{\gamma}_{\text{ILM}}(y_t | a_{n_{y_{t-1}}})}  \right\rbrace \label{eq:firstorder-rnnt-decode}
\end{equation} 
\section{Experimental Setting}
\label{exp}
The experiments are conducted on 300h Switchboard-1 (SWB) Release 2 (LDC97S62) \cite{godfrey1992SWB} and 960h LibriSpeech (LBS) \cite{povey2015librispeech}.
The evaluations for the SWB task are performed on SWB and CallHome subsets of Hub5`00 (LDC2002S09) and three subsets of Hub5`01 (LDC2002S13).
For LBS we report WERs on dev and test sets.
\ifinterspeechfinal
For training we utilize the toolkit RETURNN~\cite{doetsch2017returnn}.\ Decoding of HMM based models use RASR for the core algorithms, and its recent extension for CTC and mRNN-T decoding~\cite{rybach2011rasr,zhou2023rasr2}. Our experimental workflow is managed by Sisyphus~\cite{peter2018sisyphus}.
\else
For training and decoding we utilize two toolkits \cite{zeyer2018:returnnblind,wiesler2014rasrblind}.
\fi
The speech signal is extracted from a $25$ms window with a $10$ms shift resulting in (SWB:~40, LBS:~50) dimensional Gammatone filterbank features~\cite{schluter2007Gammatone}.\  SpecAugment is applied to all models~\cite{park2019specaugment}.\ All first-order models utilize existing setups for Conformer encoder \cite{raissi2023competitive,gulati2020conformer}.\ The model size of our 12-layers Conformer encoders is $\sim$75M parameters.\ All alignment models utilize a recurrent encoder following a similar setup to one of our previous works~\cite{raissi2022fh}.\ The recurrent encoder consists of 6 BLSTM layers with 512 nodes per direction, having $\sim$46M parameters.\
The decision to use a recurrent encoder with fewer parameters rather than the Conformer encoder for the alignment model stems from earlier experimental results indicating comparable performance of subsequent first-order models trained on their alignments~\cite{Zhou+Michel+:2022}.\  
We use one cycle learning rate schedule~(OCLR) with a peak LR of around (BLSTM: 4e-4, Conformer: 8e-4) over 90\% of the training epochs, followed by a linear decrease to 1e-6~\cite{smith2019super,Zhou+Michel+:2022}.\ The sequences are chunked into (Conformer: 400, BLSTM: 64) frames with a shift of (Conformer: 200, BLSTM: 64).\ An Adam optimizer with Nesterov momentum, together with optimizer epsilon of 1e-8 are used~\cite{dozat2016incorporating}.\ The Conformer and BLSTM models are trained for (SWB:~50, LBS:~15) and (SWB:~50, LBS:~20) epochs, respectively. We use a single consumer GPU for 5 to 15 days depending on the setup.
The reference alignments used for calculation of TSE are a triphone GMM~\cite{luscher2019rwth} for LBS and a tandem GMM~\cite{tuske2015asru} for SWB, both with speaker adaptation.\
Decoding follows a time-synchronous beam decoding with dynamic programming principle using lexical prefix tree.\ All decoding experiments use the official 4-gram language model provided with the respective tasks.\ For the real time factor measurement experiment, we used an AMD CPU (released 2021), with 2 logical cores.\
%
For further details on training hyper parameters and decoding settings, we refer to an example of our configuration setups\footnote{\ifinterspeechfinal \scriptsize{ \url{ https://github.com/rwth-i6/returnn-experiments}}\else BLIND\fi}.

\begin{table}[t]
	\setlength{\tabcolsep}{0.15em}\renewcommand{\arraystretch}{0.99}  
	\centering \footnotesize
	\caption{Comparison for LBS 960h between discriminative models with CTC and HMM topologies, using various label posterior and transition scales, $\alpha$ and $\beta$, respectively.\ In addition to the WER using a 4-gram LM, we also show the time stamp error~(TSE) of their alignment with respect to a GMM alignment, the percentage of silence~(Si) in HMM and blank~(B) in CTC, as well as the average phoneme duration(Phon). }
	\label{tab:LBS-0rder-am}
	\begin{tabular}{|c|c|c||c|c|c||c|c|}
		\hline
		\multirow{2}{*}{Model} & \multirow{2}{*}{$\alpha$} &\multirow{2}{*}{$\beta$} & \multicolumn{3}{c|}{Align model on train 960h}                                                                                                        & \multicolumn{2}{c|}{WER {[}\%{]}}                                                            \\ \cline{4-8} 
		&                  &                    & \multirow{1}{*}{TSE {[}ms{]}} & \multirow{1}{*}{Si/B {[}\%{]}} & \multirow{1}{*}{Phon.[ms]} & \multicolumn{1}{c|}{dev-other}                                & \multicolumn{1}{c|}{test-other}          \\ \hline \hline 
		GMM & 1.0 & 1.0 &  \phantom{00}0& 17.5&   \phantom{0}85.0& 19.8& - \\ \hline \hline
		\multirow{9}{*}{P-HMM} & 0.1 & \multirow{5}{*}{0.1} & 311& 60.3& \phantom{0}40.0&12.5 & 12.2 \\ \cline{2-2} \cline{4-8}
		&0.3 &  & 189& 53.7&  \phantom{0}47.5& \phantom{0}9.5 & 10.0 \\  \cline{2-2} \cline{4-8}
		&0.5 &  & 114& \phantom{0}3.4& 100.4& \phantom{0}9.2&\phantom{0}9.1 \\ \cline{2-2} \cline{4-8}
		&0.7 &  &\textbf{\phantom{0}48}& \textbf{15.1} &  \textbf{\phantom{0}88.2} & \textbf{\phantom{0}8.3}& \textbf{\phantom{0}8.8} \\  \cline{2-2} \cline{4-8}
		&1.0 &  & \phantom{0}77& 48.1& \phantom{0}53.9 & \phantom{0}8.8& \phantom{0}9.3\\  \cline{2-8}
		&\multirow{4}{*}{0.7}& 0.0 &\phantom{0}85& 46.5&\phantom{0}55.0 & \phantom{0}8.4& \phantom{0}8.6 \\  \cline{3-8}		 
		& &  0.3& \phantom{0}47& 27.1 & \phantom{0}75.8&\phantom{0}8.5 & \phantom{0}8.9\\  \cline{3-8}
		& &  0.5& 101& 55.5 &  \phantom{0}45.9& \phantom{0}9.2& \phantom{0}9.8\\ \cline{3-8}
		& &  1.0& 176 & 58.5& \phantom{0}43.0 &15.0 & 15.8\\ \hline \hline	
		\multirow{5}{*}{CTC}&0.1 & \multirow{6}{*}{ - }  &345& \phantom{0}0.2&103.0 &  26.0& 26.3 \\\cline{2-2} \cline{4-8}
		&0.3 && 240 & \phantom{0}0.5& 102.2 & 13.4& 13.5\\\cline{2-2} \cline{4-8}
		&0.5 & &180& \phantom{0}1.6& 100.9& 11.2& 11.6\\\cline{2-2} \cline{4-8}
		&0.7&&116& \phantom{0}6.1&\phantom{0}96.6& \phantom{0}9.5 & \phantom{0}9.7 \\ \cline{2-2}\cline{4-8}		
		&1.0& &\textbf{\phantom{0}38}&\textbf{54.2}& \phantom{0}\textbf{47.5}& \textbf{\phantom{0}8.0}& \textbf{\phantom{0}8.7}\\ \hline		
		
	\end{tabular}
\end{table}
\begin{table}[t]
	\setlength{\tabcolsep}{0.19em}\renewcommand{\arraystretch}{0.99}  
	\centering \footnotesize
	\caption{Similar experiments as explained in \cref{tab:LBS-0rder-am} for a subset of models using both dev and test sets for SWB 300h.}
	\label{tab:SWB-0rder-am}
	\begin{tabular}{|c|c|c||c|c|c||c|c|}
		\hline
		\multirow{2}{*}{Model} & \multirow{2}{*}{$\alpha$} &\multirow{2}{*}{$\beta$} & \multicolumn{3}{c|}{Align model on train 300h}                                                                                                        & \multicolumn{2}{c|}{WER {[}\%{]}}                                                            \\ \cline{4-8} 
		&                  &                    & \multirow{1}{*}{TSE {[}ms{]}} & \multirow{1}{*}{Si/B {[}\%{]}}& \multirow{1}{*}{Phon.[ms]} & \multicolumn{1}{c|}{hub5`00}                                & \multicolumn{1}{c|}{hub5`01}          \\ \hline \hline
		GMM &  1.0& 1.0 &  \phantom{00}0 & 25.1& \phantom{0}86.5 & 18.9& - \\ \hline \hline
		\multirow{4}{*}{P-HMM} & 1.0&  \multirow{2}{*}{0.1} & \phantom{0}79& 50.4 &  \phantom{0}57.2& 14.9&14.1 \\ \cline{2-2} \cline{4-8}
		
		&\multirow{2}{*}{0.7}&  &\textbf{\phantom{0}62} & \textbf{27.2}& \textbf{\phantom{0}84.0} & \textbf{13.7}&\textbf{13.0} \\ \cline{3-8}
		&&0.0 & \phantom{0}71 &20.5&  \phantom{0}91.7& 14.0& 13.4\\\cline{2-8}
		&0.5 &  0.1 & 117& 17.4&  \phantom{0}95.3&14.7 & 14.3\\  \hline \hline
		\multirow{3}{*}{CTC} &1.0 & \multirow{3}{*}{-} &  \textbf{\phantom{0}58}& \textbf{54.9} &\textbf{\phantom{0}51.9} & \textbf{13.7}& \textbf{13.0} \\ \cline{2-2}  \cline{4-8} 
		&0.7 & &129 & 19.2&\phantom{0}93.3 &  14.2& 13.7\\ \cline{2-2}  \cline{4-8}		
		& 0.5& &240& 12.3 &101.1 & 16.7& 16.6\\ \hline
	\end{tabular}\vspace{-0.4cm}
\end{table}

\begin{table}[t]
	\setlength{\tabcolsep}{0.2em}\renewcommand{\arraystretch}{0.99}  
	\centering \footnotesize 
		\caption{ ASR performance of factored hybrid HMM~(FH) and mRNN-T in terms of WER for LBS960 and SWB300h, on the alignments presented in \cref{tab:LBS-0rder-am,tab:SWB-0rder-am}.} 
		\label{tab:LBS-1order-am}
		\begin{tabular}{|c||c|c|c|c||c|c|}
			\hline
			\multirow{3}{*}{Model}  & \multicolumn{4}{c|}{Align Model on LBS 960} & \multicolumn{2}{c|}{WER {[}\%{]}}  \\ \cline{2-7} 
			& $\alpha$& $\beta$& TSE {[}ms{]} & Si/B [\%]   & \multicolumn{1}{c|}{dev-other}            &                  \multicolumn{1}{c|}{test-other}          \\ \hline
			\multirow{3}{*}{FH} & 1.0 & \multirow{3}{*}{0.1}&\phantom{0}77&48.1&\phantom{0}7.0& \phantom{0}7.6  \\ \cline{2-2} \cline{4-7}
			&  0.7&&\textbf{\phantom{0}48}&\textbf{15.1}&\textbf{\phantom{0}6.4}& \textbf{\phantom{0}6.9}  \\ \cline{2-2} \cline{4-7}
			&0.5 &&114&\phantom{0}3.4&\phantom{0}6.8& \phantom{0}7.2 \\ \hline \hline
			\multirow{3}{*}{mRNN-T}&1.0&\multirow{3}{*}{-}&\textbf{\phantom{0}40}&\textbf{61.3}&\textbf{\phantom{0}6.8}&  \textbf{\phantom{0}7.1} \\ \cline{2-2} \cline{4-7}
			&0.7 & &610&11.4&\phantom{0}7.5& \phantom{0}7.9 \\ \cline{2-2}  \cline{4-7}			
			&0.5 &&609&17.8&20.0&22.3\\ \hline
		\end{tabular} 
	\end{table}
		\begin{table}[t]
				\setlength{\tabcolsep}{0.25em}\renewcommand{\arraystretch}{0.99}  
			\centering \footnotesize 
			\caption{ Similar experiments explained in \cref{tab:LBS-1order-am} for SWB300.} 
			\label{tab:SWB-1order-am}
		\begin{tabular}{|c||c|c|c|c||c|c|}
			\hline
			\multirow{3}{*}{Model}  & \multicolumn{4}{c|}{Align Model on SWB 300h} & \multicolumn{2}{c|}{WER {[}\%{]}}  \\ \cline{2-7} 
			& $\alpha$& $\beta$ &TSE [ms] & Si/B [\%]  & \multicolumn{1}{c|}{hub5`00}  &   \multicolumn{1}{c|}{hub5`01}          \\ \hline
			\multirow{3}{*}{FH} & 1.0 & \multirow{3}{*}{0.1}&\phantom{0}79&50.4&11.7  &11.6   \\ \cline{2-2} \cline{4-7}
			&0.7 & &\textbf{\phantom{0}62}&\textbf{27.2}&\textbf{11.4}& \textbf{11.2}\\  \cline{2-2} \cline{4-7}
			& 0.5&& 117&17.4&11.9& 11.5 \\ \hline \hline
			\multirow{3}{*}{mRNN-T}&1.0 &\multirow{3}{*}{-} &\textbf{\phantom{0}62}&\textbf{65.3}&\textbf{12.5}& \textbf{12.2}\\ \cline{2-2}  \cline{4-7}
			&0.7& &127&65.2&13.4& 12.3\\ \cline{2-2}  \cline{4-7}
			&0.5 &&236&65.3&14.4&13.7\\ \hline
		\end{tabular}\vspace{-0.2cm}
\end{table}
\vspace{-0.2cm}
\section{Evaluation}
\subsection{Alignment Quality}
\label{sec:alignquality}

Our alignment models of \cref{subsub:zeroorder} are trained from scratch with the full-sum training criterion of \cref{eq:fsloss}.\ A CUDA implementation of the dynamic programming formulation of the forward-backward algorithm computes the state marginals used in the calculation of the loss~\cite{Zeyer+Beck+:Interspeech2017}.\ The kernel is agnostic to the label topology, enabling a close comparison between the two models.\

The P-HMM formulated in \cref{eq:HMM} consists of a label posterior and a transition probability.\ By scaling each component, one can control their distribution shape and therefore their frame-level log-linear contribution.\ This has an effect on the certainty with which the model would choose to stay in a label segment or emit a new label.\ Therefore, there is a close relation between the input frame shift (classic 10ms against our 40ms), and the choice of aforementioned scales.\ Viewing the training as a general expectation-maximization~(EM) procedure, the scales are then applied before passing the scores to the alignment FSA, i.e. at the expectation step, guaranteeing the local normalization constraint at the maximization step.\ 

Differently to the WER that is a well-defined evaluation metric, it is not clear how to evaluate the quality of the alignment of our models.\ Due to the lack of a proper ground-truth, a set of different measurements in our case is considered.\ In addition to the WER of the alignment model, we consider the time stamp error~(TSE), i.e. the mean absolute distance (in milliseconds) of word start and end positions against a reference GMM alignment, irrespective of the silence~\cite{Raissi+Zhou+:2023,zhang2021lattice}.\ Furthermore, we include the average phoneme duration, as well as the percentage of silence and blank frames for HMM and CTC alignments, respectively.\ We then choose within each topology the model with best combination of TSE and WER, and train a first-order model using its alignment.\ As a final measure, we consider the WER of this system, reported in \cref{sec:asrperformance}.\ There are different aspects to be taken into account when comparing the statistics and measures between the two models.\ Due to the ambiguous role of the blank label in CTC topology, the average phoneme duration is not comparable between the two models, since part of the phoneme duration might be consumed by blank in CTC.\ This applies also to the comparison between silence and blank percentage.\ Furthermore, when transferring the CTC alignment for RNN-T training, label loops are removed, following the topology.\ For the statistics reported in \cref{tab:LBS-0rder-am,tab:SWB-0rder-am}, we use the alignment without the mentioned post-processing, and we will report the TSE of the post-processed alignment in \cref{sec:asrperformance}.\ 
 The results show that differently to CTC and its peaky alignment, the P-HMM with lowest WER has a phoneme duration and silence percentage that is near to the reference GMM alignment.\ In the CTC alignment, we observed that the word-starts tend to be shifted forward while the word-ends tend to be shifted backward in comparison to the reference.\ This may indicate that the label peaks are typically placed around the center of the actual label duration.\ 

We examined different combinations of scales for label posterior and transition probability in HMM, along with the label posterior in CTC.\ The results indicated that the optimal scales of each topology were task-independent, i.e. the same for LBS and SWB.\ Unsurprisingly, CTC necessitates no scaling of the label posterior, i.e. the optimal value is $1.0$.\ For HMM, the optimal label posterior and transition scales were found to be $0.7$ and $0.1$, respectively.\ The optimal label posterior scale for a P-HMM trained with 10ms shift, was shown to be $0.3$~\cite{Raissi+Zhou+:2023}.\ This confirms the relation between the scale and the frame shift.\ To further verify whether the encoder output with 40ms has less correlated input information at each step, we also use the acoustic lookahead temporal approximation scale~(ALTAS) during decoding~\cite{nolden2013advanced}.\ In this method the acoustic lookahead score for time $t+1$ is approximated by the scaled score at time $t$.\ This pruning method works well, if the scores between adjacent frames are highly correlated.\ Therefore, increasing the ALTAS value without degrading WER suggests a stronger correlation among adjacent frames.\ The faster degradation of the 40ms P-HMM can be seen in \cref{fig:altas-wer}.\ 
\begin{figure}[t]
	\centering
	\includegraphics[width=0.65\columnwidth, trim={0.0cm 0cm 0cm 0cm},clip]{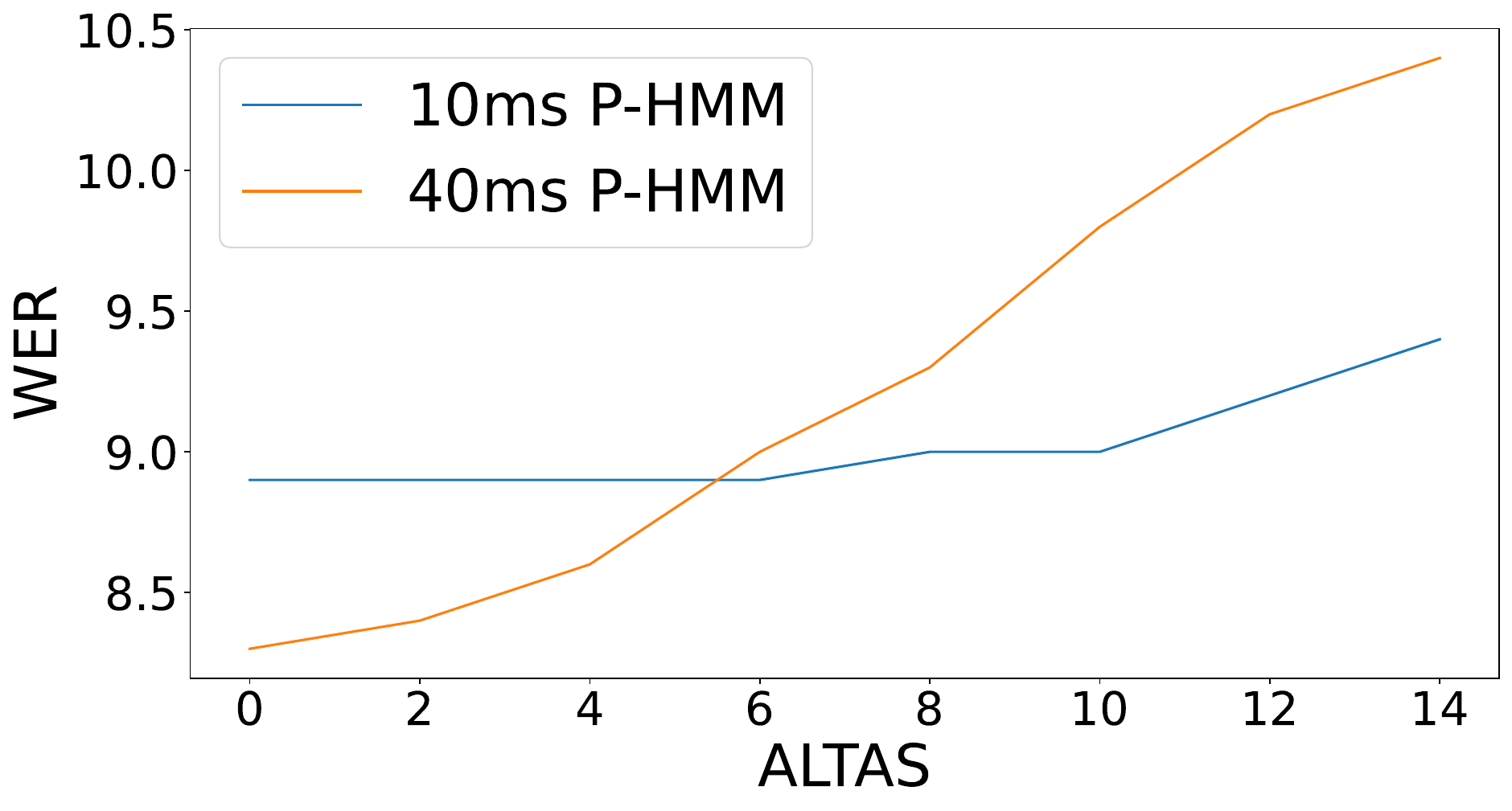}
	\caption{comparison of WER on dev-other between two P-HMMs with 10ms and 40ms frame shift when using acoustic lookahead temporal approximation scale (ALTAS).\ Note that the scale value represents a different temporal range for each curve.}
	\label{fig:altas-wer} 
\end{figure} 
\begin{table}[t]	
	\setlength{\tabcolsep}{0.05em}\renewcommand{\arraystretch}{1.0}  
	\centering \footnotesize 
	\caption{Optimal real time factors~(RTF), with distinction between the encoder~(Enc.) and search for best 40ms factored hybrid~(FH) and mRNN-T models, as well as 10ms FH and traditional hybrid-HMM.\ Reporting the average number of active state~(\#S) and lexical prefix-trees~(\#L) hypotheses at each step, together with WERs and the min and max values for the confidence interval.}
	\label{tab:rtf}
		\begin{tabular}{|c|c||c|c|c|c|c|c||c|c|} 
			\hline			
			 \multirow{2}{*}{Model} & Frame&  \multirow{2}{*}{\#S} & \multirow{2}{*}{\#L}& \multicolumn{3}{c|}{RTF} & WER[\%] & \multicolumn{2}{c|}{Conf. Int.}\\ \cline{5-10}
			 & Shift[ms] & && Search& Enc. & $\scaleto{\sum}{6pt}$& dev-other &min & max\\ 
			  \hline \hline
			Hybrid-HMM & \multirow{2}{*}{10} & 120&\phantom{0}14&0.02&0.15&0.17& 6.6 &5.2& 8.4\\ \cline{1-1} \cline{3-10}
			\multirow{2}{*}{FH} &	&410& \phantom{0}56 & 0.04&0.21 &0.25&6.5&5.2& 8.2\\  \cline{2-10}
			&\multirow{2}{*}{40}   &\phantom{0}47 &\phantom{0}71 &0.01&\multirow{2}{*}{0.10}&0.11&6.4&5.2& 8.0\\  \cline{1-1}\cline{3-5} \cline{7-10}
			mRNN-T				    & & \phantom{0}11& \phantom{00}4&  0.05& &0.15&6.8&5.5&8.4 \\ \hline 
		\end{tabular}   \vspace{-0.5cm}

\end{table}

\subsection{ASR Performance}
\label{sec:asrperformance}
In \cref{tab:LBS-1order-am,tab:SWB-1order-am}, it is possible to verify the performance of the models, using the best alignment for each topology from \cref{sec:alignquality}.\ Even though the CTC alignments had slightly better WER and TSE, here the subsequent training favors the HMM approach on both tasks.\ Finally, we compare the decoding real time factor using a simulated production setup with 75\% load on the machine.\ It is possible to see that the 40ms shift brings up to 50\% relative speedup to the FH, surpassing also the classic hybrid-HMM.\ 

\vspace{-0.5cm}
\section{Conclusions}
In this work, we compared two context-independent alignment models with CTC and HMM topologies, and two first-order label context models with monotonic RNN-T and factored HMM topologies.\ We examined the alignment quality by using various measurements and compared the ASR performance  and the real time factor of our best systems during decoding.\ Moreover, we showed that also the HMM topology can benefit from larger frame shift in the neural encoder architectures, respecting its inherent output independence assumption.

\section{Acknowledgements}
This work was partially supported by NeuroSys, which as part of the
initiative ''Clusters4Future'' is funded by the Federal Ministry of
Education and Research BMBF (03ZU1106DA). Sincere appreciation is extended to Peter Bell (University of Edinburgh) and Lukáš Burget (Brno University of Technology) for insightful conversations and invaluable comments. Authors thank Noureldin Bayoumi and Moritz Gunz for training the CTC and the CART hybrid models, respectively.
\bibliographystyle{IEEEtran}
\bibliography{references}

\end{document}